\documentclass{osa-article}
\journal{oe}
\articletype{Research Article}
\usepackage{lineno}

\usepackage{subfig, gensymb}
\usepackage{bm}
\usepackage{amsmath}
\usepackage{color}
\usepackage[normalem]{ulem} 

\begin{document}

\title{{Sensing the position of a single scatterer in an opaque medium by mutual scattering}}

\author{Minh Duy Truong,\authormark{1} 
Ad Lagendijk,\authormark{1} and 
Willem L. Vos\authormark{1,*}}

\address{\authormark{1}Complex Photonic Systems (COPS), MESA + Institute for Nanotechnology, University of Twente, P.O. Box 217, 7500 AE Enschede, The Netherlands}

\email{\authormark{*} Email: w.l.vos@utwente.nl} 

\homepage{Website: https://nano-cops.com/} 


\begin{abstract}
We investigate the potential of mutual scattering, i.e., light scattering with multiple properly phased incident beams, as a method to extract structural information from inside an opaque object. 
In particular, we study how sensitively the displacement of a single scatterer is detected in an optically dense sample of many (up to $N=1000$) similar scatterers. 
By performing exact calculations on ensembles of many point scatterers, we compare the mutual scattering (from two beams) and the well-known differential cross-section (from one beam) in response to the change of location of a single dipole inside a configuration of randomly distributed similar dipoles. 
Our numerical examples show that mutual scattering provides speckle patterns with an angular sensitivity at least 10 times higher than the traditional one-beam techniques.
By studying the ``susceptivity'' of mutual scattering, we demonstrate the possibility to determine the original depth relative to the incident surface of the displaced dipole in an opaque sample. 
Furthermore, we show that mutual scattering offers a new approach to determine the complex scattering amplitude.
\end{abstract}


\section{Introduction}

Traditional scattering experiments are associated with sending a single beam of waves to a target~\cite{Warren1969XrayDif, Ishimaru1978Book, AlsNielsen2001Book, Chu2007Book}. 
If the target interacts only weakly with the incoming waves, as is typical for X-ray and light scattering~\cite{Warren1969XrayDif, Ishimaru1978Book, AlsNielsen2001Book, Chu2007Book}, the detailed structure of the target including the positions of constituting particles can be retrieved from the characteristics of the scattered waves. 
If a medium interacts more strongly with the waves~\cite{Vos1996PRB, Vos2015Book}, it becomes increasingly opaque, hence the usual single scattering approaches break down, and only limited structural information can be retrieved by methods such as diffusion wave spectroscopy~\cite{Pine1988PhyRevLet, Maret1987PhyBConMat, Stetefeld2016BioRev}. 
Beyond the traditional case of a single incident wave~\cite{Warren1969XrayDif, Ishimaru1978Book, AlsNielsen2001Book, Chu2007Book}, the recent development of multiple-beam techniques such as wavefront shaping has opened the new potential for research of strongly interacting opaque samples~\cite{Mosk2012NatPhot, Rotter2017RevModPhy, Cao2022NatPhys}. 

Recently, it was realized that interference of \textit{multiple} incident beams gives rise to a new scattering phenomenon, called mutual scattering~\cite{Lagendijk2020EPL}. 
From a generalized optical theorem, it follows that the total extinction of even only two incident waves is either greatly enhanced by up to $100\%$ compared to the usual single-beam extinction, which is called mutual extinction, or greatly reduced, which is called mutual transparency. 
Subsequently, the experimental demonstration of mutual scattering has been made for biological, silicon, and carbon samples~\cite{Rates2021PhyRevA}.

Following the theoretical and experimental demonstration, we explore here a new practical application of mutual scattering, namely: how sensitive is mutual scattering to detecting the displacement of a single nanoparticle deep inside a sample with an ensemble of many (up to 1000) similar nanoparticles? 
In other words, the nanoparticle we wish to track is not a tracer particle with properties different from the ensemble, like a dyed nanosphere standing out in a sea of undyed ones~\cite{Hong2018Optica}. 
The underlying hypothesis is that the cross-interference of two beams is more sensitive to changes of a scatterer located deep within the sample than conventional scattering methods. 
An enhanced sensitivity would boost the potential of mutual scattering for applications in biomedical imaging inside tissue, in semiconductor metrology, in communications for non-line-of-sight links~\cite{Raptis2016JouOptSocAmeA}, or in the study of the structure of free-form samples~\cite{Haberko2013PhyRevA,Falaggis2022OptExp}. 

In addition, we will also show that mutual scattering allows one to extract the modulus and imaginary part of the scattering amplitude in an experiment. 
By combining this new information with the traditional one-beam scattering, we obtain the complex scattering amplitude \textit{at all angles}, while this was up to now only possible for forward scattering~\cite{Moteki2021OptExp}. 
In other words, mutual scattering of multiple beams is an interferometric technique to measure both the amplitude and phase of the complex scattering amplitudes.

\section{Schematic of mutual scattering}
\label{sec2}
\begin{figure}[ht!]
\centering
\includegraphics[width=0.55\textwidth]{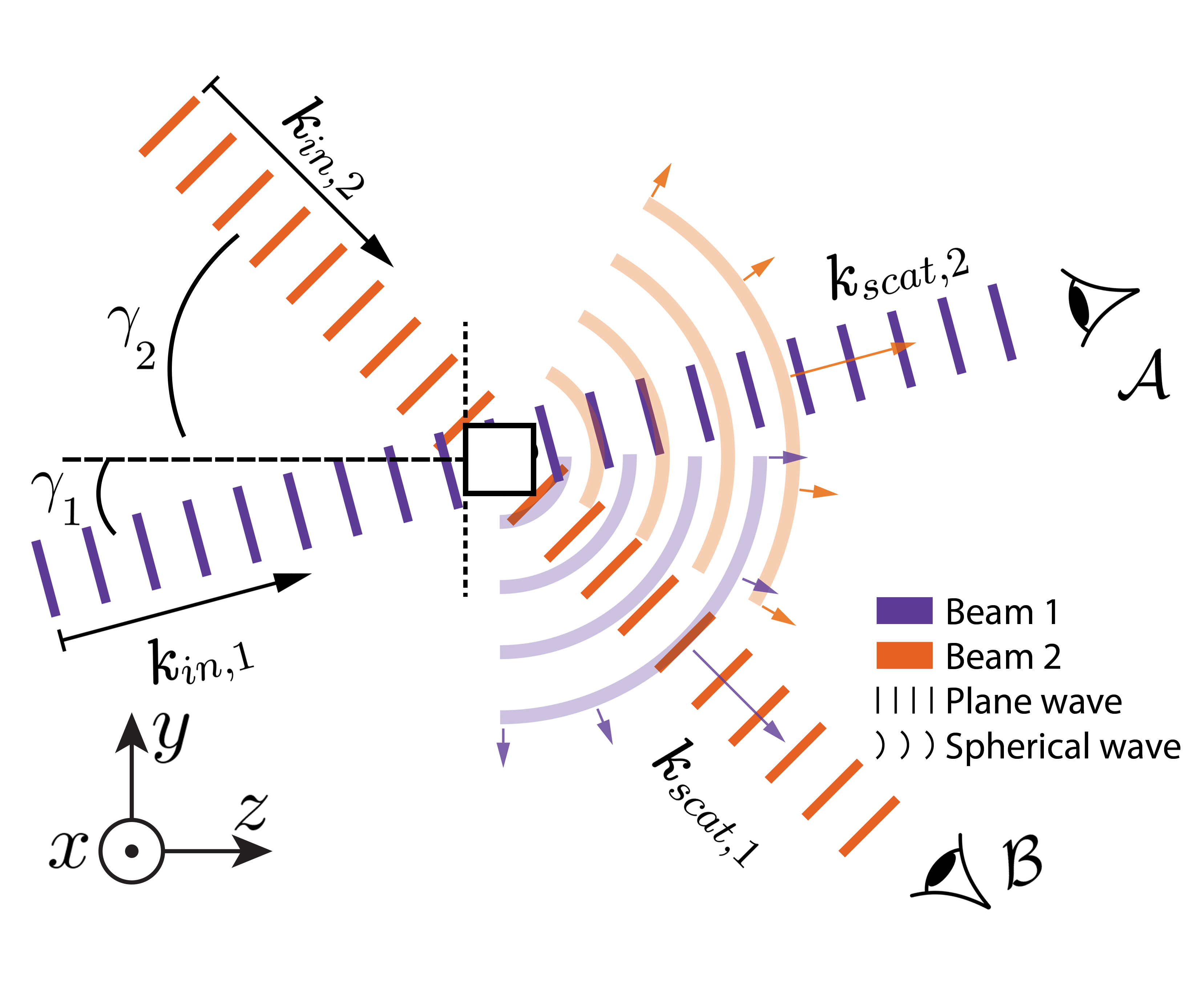} 
\caption{\label{fig:geo1}
Schematic of mutual scattering:
the sample (white box) at the centre is illuminated from the left by two incident (purple and orange) plane waves. 
The interference between the incident plane waves and the scattered spherical waves of different colors, observed at detectors $\mathcal{A}$ or $\mathcal{B}$, gives rise to the mutual scattering phenomena. 
}
\end{figure}

To study the characteristics of mutual scattering, we introduce the sample geometry consisting of a box-like sample as shown in Fig.~\ref{fig:geo1}. 
The sample has thus finite support in three dimensions (3D)~\cite{Hasan2018PhyRevLet}. 
We illuminate the left side of the sample with two incident beams with wave vectors $\mathbf{k}_{\text{in},1}$ and $\mathbf{k}_{\text{in},2}$, respectively, and with equal amplitudes $A_1 = A_2 = A$. 
The incident plane is chosen to be perpendicular to the $x$-axis. 
The angles of incidence of these two beams, denoted $\gamma_1$ and $\gamma_2$, respectively, are defined as the angles between the $z$-axis and $\mathbf{k}_{\text{in},1}$ or $\mathbf{k}_{\text{in},2}$. 
The two incident beams subtend an angle $\gamma = \gamma_1 + \gamma_2$. 
After emanating from the right of the sample and experiencing a certain mutual scattering, the two beams propagate to two detectors located at $\mathcal{A}$ and $\mathcal{B}$. 
Detector $\mathcal{A}$ measures the mutual scattering $F^\text{MS}_1(\gamma)$ along $\mathbf{k}_{\text{in},1}$, and detector $\mathcal{B}$ measures $F^\text{MS}_2(\gamma)$, see Appendix~\ref{appendix} for details. 

\begin{figure}[ht!]
\centering
\includegraphics[width=0.78\textwidth]{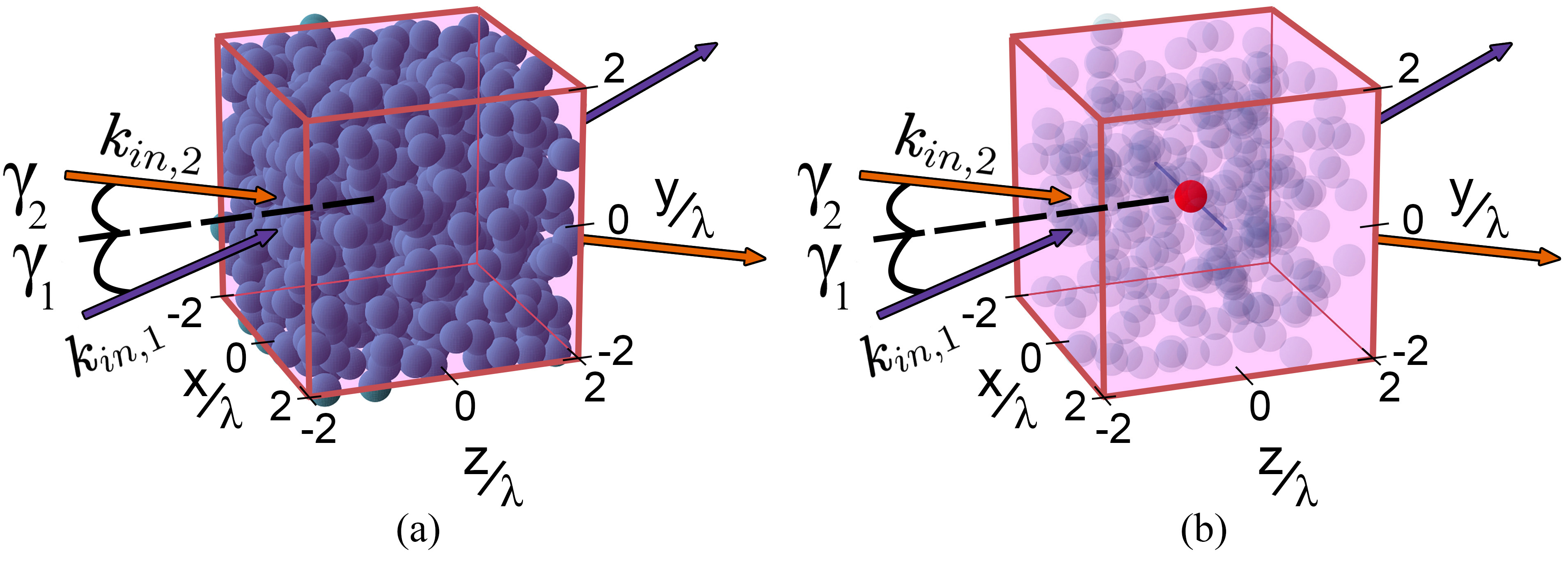}
\caption{\label{fig:sche2} 
Schematic of the numerical samples: 
(a) A cube with $N_\text{dipole}=1000$ dipoles, and (b) a cube with $N_\text{dipole}=250$ dipoles, whose polarizability is shown by the extent of the blue spheres. 
The target dipole (red sphere) at position $\mathbf{r}_0$ is moved along a chosen direction, for example, the blue line shows the movement of the red scatterer in the $x$-direction, while the positions of all other ($N_\text{dipole}-1$) scatterers are preserved. 
}
\end{figure}

To model the optical properties of opaque objects, we consider an ensemble of $N_\text{dipole}$ scatterers, shown in Fig.~\ref{fig:sche2}(a,b), that are randomly distributed in a rectangular box with a volume $V = 4\times 4\times 4$ $\lambda^3$, with $\lambda$ the wavelength of light. 
The scatterers are dipoles, whose optical properties are described in Appendix~\ref{sub:Tmatrix}, where their interaction strength with light~\cite{Vos1996PRB, Vos2015Book} is illustrated by the volume of the polarizability spheres in Figs.~\ref{fig:sche2}(a,b). 
The sample in Fig.~\ref{fig:sche2}(a) holds $N = 1000$ scatterers and has a substantial photonic strength that corresponds to a mean free path less than the wavelength $l_\text{scat}/\lambda = 0.2011$ or less than the sample dimensions $V^{1/3}/l_\text{scat} = 20$, typical of a highly opaque sample. 
The sample in Fig.~\ref{fig:sche2}(b) interacts less strongly with light, as it has $l_\text{scat}/\lambda = 0.8044$ or $V^{1/3}/l_\text{scat} = 5$ and is thus still opaque. 

To compute the scattering amplitude of the sample, 
we must compute the exact evaluation of the T-matrix of many scatterers, see Appendix \ref{sub:Tmatrix} for more details. 
The scattering amplitude $f$ is obtained from
\begin{align}\label{eq:tmatrix}
f(\mathbf{k}_{\text{out}}, \mathbf{k}_{\text{in}}) = - \dfrac{1}{4\pi} T(\mathbf{k}_{\text{out}}, \mathbf{k}_{\text{in}}),
\end{align}
where $T(\mathbf{k}_{\text{out}}, \mathbf{k}_{\text{in}})$ is the transition matrix or T-matrix in scattering theory~\cite{Lagendijk1996PhyRep}. 

We note aside that when the sample is opaque and significantly larger than the wavelength, the scattering amplitude is accurately described with Fraunhofer diffraction theory~\cite{Born1999PrinciplesOptics}, as shown in Ref.~\cite{Rates2021PhyRevA}. 
However, since the sample is considered to be impenetrable in this description, this approximation is not helpful to study the internal structure. 

\section{Results I - Mutual scattering of static configuration}\label{sec:static}
We study the properties of mutual scattering in response to varying the static configuration of the sample. 
We start with an ensemble of $N_\text{dipole}= 250$ dipoles randomly distributed in a rectangular box and create new configurations by randomly changing the positions of all dipoles. 

\subsection{Symmetric incident beams}
Based on the theory in Eq.~(\ref{eq:Jext1}), we find that the mutual scattering depends on the phase difference $(\phi_1-\phi_2)$ between two beams. 
By tuning the phase difference, we obtain the maximum and minimum mutual scattering for each angle of incidence $\gamma$ between the two incident beams. 

We start with the symmetric case, where the two incident beams with mutual angle $\gamma$ symmetrically illuminate the sample, such that the angles of incidence of both beam 1 and beam 2 are equal to $\gamma/2$. 
For each sample configuration, the mutual scattering amplitudes $F^\text{MS}_1$ and $F^\text{MS}_2$ are not equal because a single random structure is always asymmetric. 
If we average the mutual scattering over many configurations, however, the statistics of the mutual scattering of beam 1 and beam 2 and the mutual scattering of both beams will be the same:
\begin{align}
\langle F^\text{MS}_{1} \rangle= \langle F^\text{MS}_{2} \rangle = \langle F^\text{MS}_{12} \rangle,
\end{align}
where the notation $\langle x \rangle$ represents the average (mean) value of $x$ calculated over $N_\text{realization}$ realizations. 
Therefore, we will show only the result of $F^\text{MS}_{12}$ in this symmetric case.

\begin{figure}[ht!]
\centering
\includegraphics[width=0.8\textwidth]{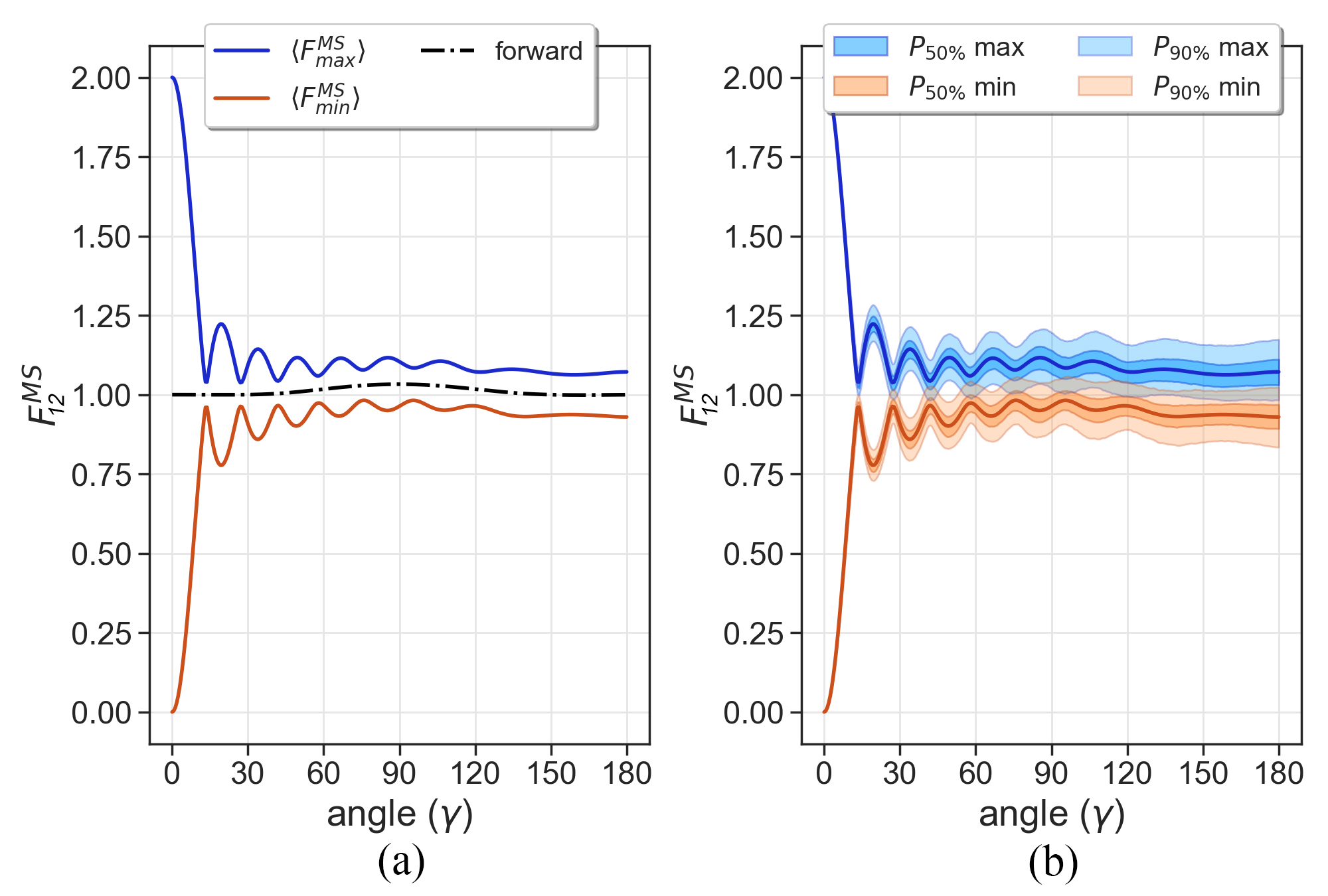}
\caption{\label{fig:statmet3} Mutual scattering $F^\text{MS}_{12}$ for $N_\text{dipole}= 250$ with respect to angle $\gamma$ between two symmetric incident beams. 
The results are averaged over $N_\text{realization}=1000$ realizations of different configurations of the location of dipoles. 
(a) The blue and orange lines correspond to the maximum and minimum scattering obtained from phase variations, respectively. 
The average self-extinction containing only the sum of the forward scattering is given by the black dotted line. 
(b) The percentile distribution of the maximum and minimum mutual scattering.}
\end{figure}

Fig.~\ref{fig:statmet3} shows the statistics of the maximum and minimum mutual scattering, denoted by $F^\text{MS}_\text{max}$ and $F^\text{MS}_\text{min}$, respectively, as a function of angle $\gamma$ between the two incident beams for $N_\text{dipole}= 250$. 
The results are calculated based on $N_\text{realization}=1000$ realizations of different configurations of the location of scatterers.
Fig.~\ref{fig:statmet3}(a) shows both the average value of the maximum $\langle F^\text{MS}_\text{max} \rangle$ and minimum mutual scattering $\langle F^\text{MS}_\text{min} \rangle$ over all the realizations as functions of angle $\gamma$. 
The average value of the forward-scattering 
of both beams $\langle F^\text{forward}_{12}(\gamma) \rangle$ is given by 
\begin{align}
\langle F^\text{forward}_{12}(\gamma) \rangle = \dfrac{\langle F^\text{MS}_\text{max}(\gamma) \rangle + \langle F^\text{MS}_\text{min}(\gamma) \rangle}{2}. 
\end{align}
At $\gamma=0\degree$, the forward-scattering equals 1, increases gradually and peaks at $\gamma=90\degree$, when the incident angles of two beams equal $45^{\degree}$. 
The average maximum and minimum mutual scattering, i.e., $\langle F^\text{MS}_\text{max}(\gamma) \rangle$ and $\langle F^\text{MS}_\text{min}(\gamma) \rangle$, have a shape like a sine cardinal function near $\langle F^\text{forward}_{12}(\gamma) \rangle$. 
At small angles, the mutual scattering is strong, with modulations up to 100\%. 
The mutual scattering quickly decreases with increasing angle, being up to 20\% at $\gamma = 20\degree$ and then further decreasing. 
The maximum and minimum mutual scattering vary according to the distribution of scatterers inside the box. 

The statistics of that variation are shown in Fig.~\ref{fig:statmet3}(b), notably the 50-percentile and 90-percentile distribution of both the maximum and minimum mutual scattering. 
It is apparent that the mutual scattering amplitudes vary strongly with varying configurations.
For example, at $\gamma =30\degree$, there is a $90\%$ chance out of 1000 realizations that the value of the maximum mutual scattering is found in the range between 1 and 1.2.
The variations increase with increasing angle, such that at angles in excess of $\gamma =60\degree$ the variation ranges of maximum and minimum start to overlap.

\begin{figure}[ht!]
\centering
\includegraphics[width=0.75\textwidth]{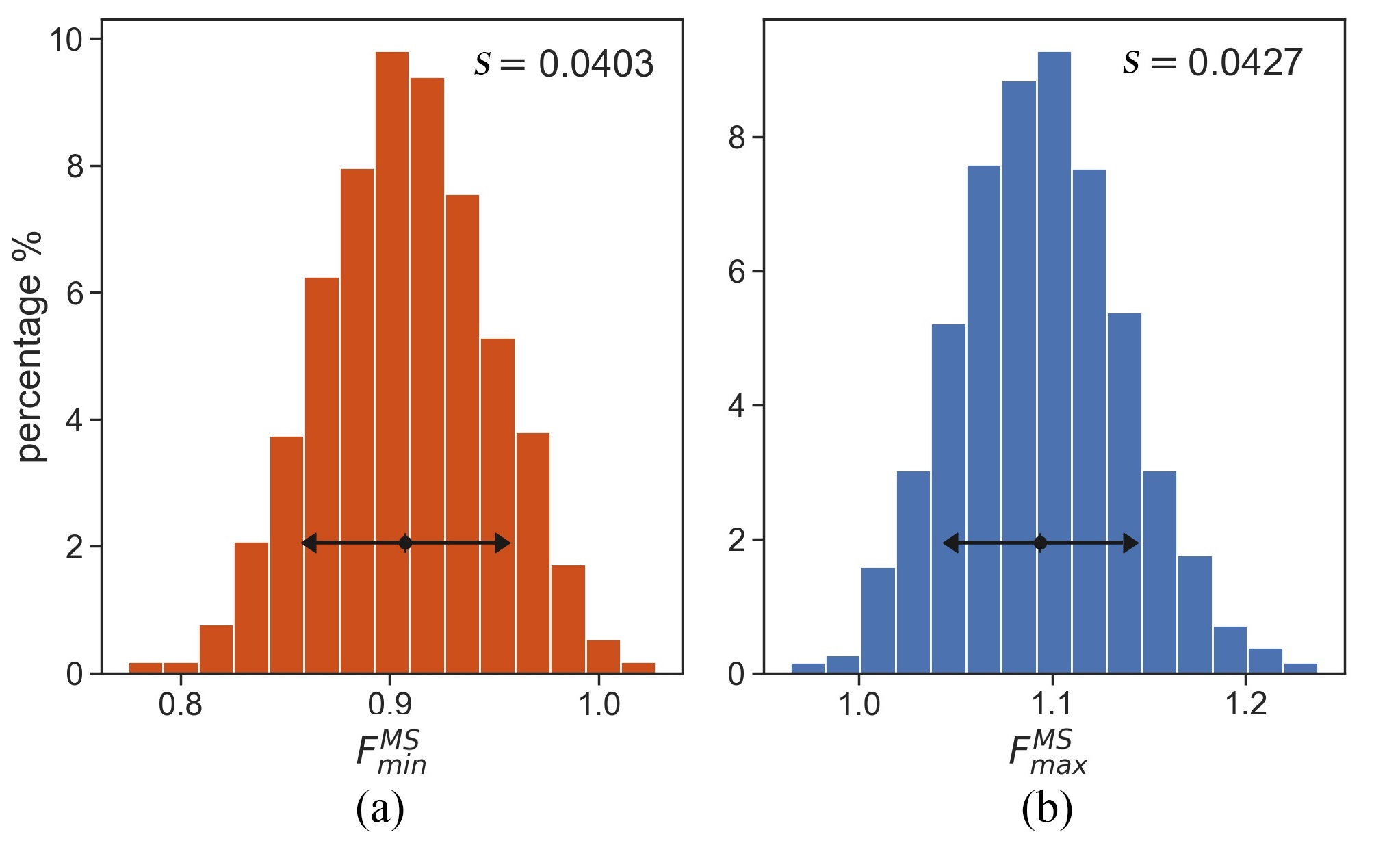}
\caption{\label{fig:statmet1c30}
15-bin percentage histogram of maximum (blue) and minimum (orange) mutual scattering of symmetric incident beams for $N_\text{dipole}= 250$ and $N_\text{realization}=1000$ at angle $\gamma=30\degree$. 
The black circle in the centre of the double-headed arrow represents the mean value, and the standard deviation $s$ of the maximum and minimum mutual scattering over all realizations is indicated by the black double-headed arrows. 
}
\end{figure}

The mutual scattering statistics are more readily apparent at a constant angle as shown in Fig.~\ref{fig:statmet1c30}(a,b) for $\gamma=30\degree$, which shows the histograms of the maxima and minima for symmetric incident beams for $N_\text{dipole}= 250$ and $N_\text{realization}=1000$. 
In this representation, it is apparent that both the maximum and minimum mutual scattering amplitudes show considerable variations over all realizations. 
In one extreme realization, where the maximum and minimum mutual scattering can both take the value 1.0, the whole mutual scattering effect is completely washed out. 
On the other hand, when averaged over all realizations, however, there is a significant mutual scattering, with the mean being 1.1 (maximum) and 0.9 (minimum), where the difference exceeds the indicated standard deviations. 
Furthermore, Fig.~\ref{fig:statmet3}(b) and Fig.~\ref{fig:statmet1c30}(a,b) show that the probability distributions of both maximum and minimum mutual scattering at a fixed angle $\gamma$ are neither symmetric nor normally distributed. The maximum scattering is right-skewed, while the minimum is left-skewed. 

\begin{figure}[ht!]
\centering
\includegraphics[width=0.75\textwidth]{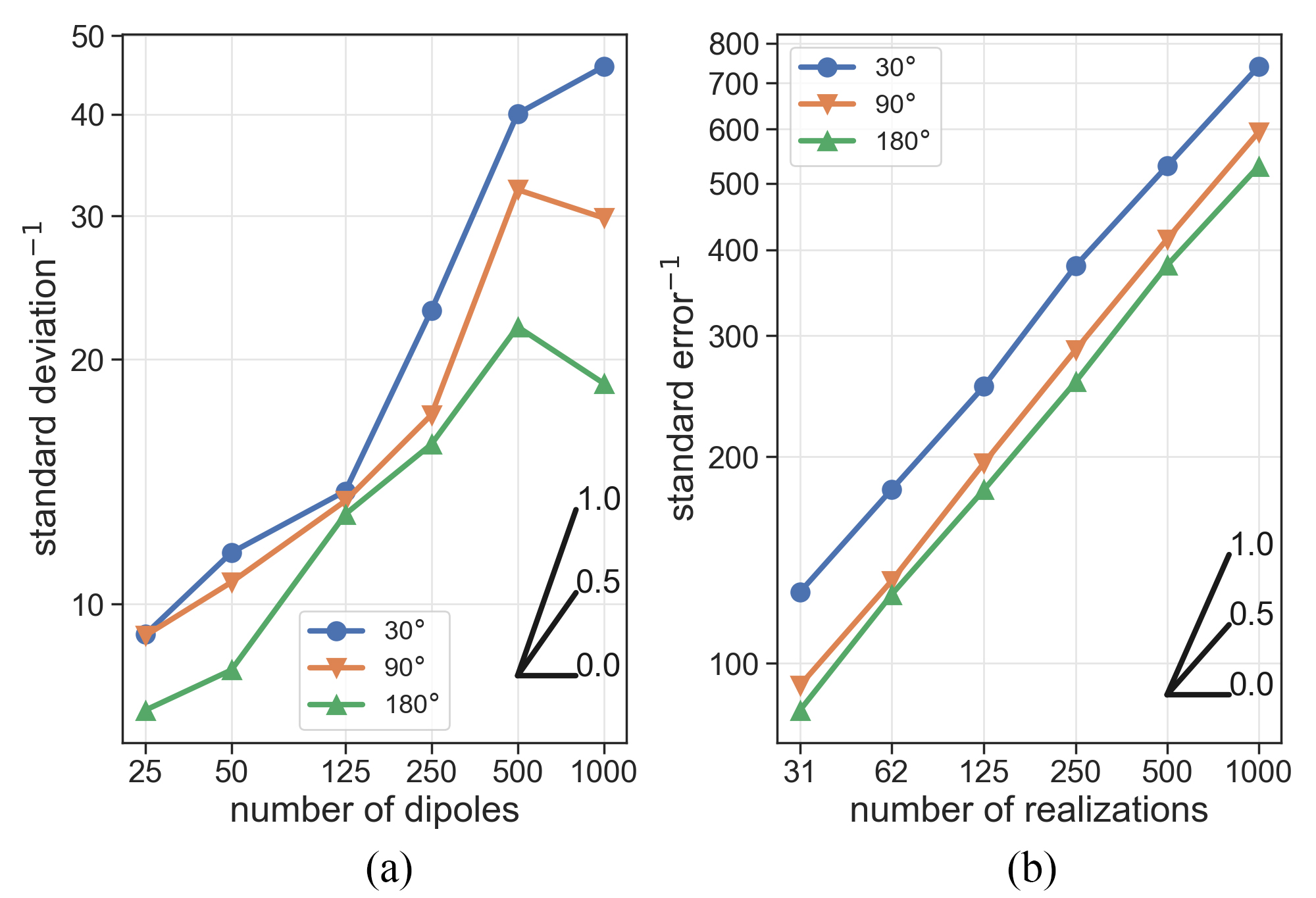}
\caption{\label{fig:dd} 
(a) Inverse standard deviation $s^{-1}$ of the distribution of the maximum mutual scattering of two symmetric incident beams versus the number of dipoles $N_\text{dipole}$. 
The results are averaged over $N_\text{realization}=50$ realizations. 
(b) Inverse standard error $\hat{s}^{-1}$ of the distribution of the maximum mutual scattering of two symmetric incident beams versus the number of realizations $N_\text{realization}$. 
The results are computed for a given number of dipole scatterers $N_\text{dipole}=250$. 
The black slopes in the bottom right corners of (a,b) indicate three different power laws ($y\propto x^k$ where $k=0, 0.5,$ and $1$). 
}
\end{figure}

Figure~\ref{fig:dd}(a) shows the inverse standard deviation as a function of the number of dipoles, and Figure~\ref{fig:dd}(b) shows the inverse standard error of the mean $\hat{s}$ versus the number of realizations, for three representative angles ($\gamma = 30\degree,  90\degree$, and $180\degree$). 
Naively, we expect that an increase in the number of dipoles, corresponding to an increase in the number of scattering events, would increase the standard deviation of mutual scattering.
However, the inverse standard deviation overall increases with the number of dipoles with power between 0.5 and 1.0 (Fig.~\ref{fig:dd}(a)). 
On the other hand, as a function of the number of realizations, all three curves of the inverse standard error $\hat{s}^{-1}\coloneqq s^{-1}{N_\text{realization}^{1/2}}$ in Fig.~\ref{fig:dd}(b) are closely consistent with a power 0.5.
This implies that the standard deviation $s$ is almost independent of the number of realizations.


The standard deviation $s$ of mutual scattering in Fig.~\ref{fig:dd}(a), and the standard error of the mean $\hat{s}$ of mutual scattering in Fig.~\ref{fig:dd}(b), increase as the angle between the two incident beams increases. 
Indeed, the blue line (for $\gamma = 30\degree$) is always above the orange line (for $\gamma = 90\degree$), which is, in turn, higher than the green line (for $\gamma = 180\degree$).  
This is consistent with Fig.~\ref{fig:statmet3}(b), where mutual scattering fluctuates with larger amplitude at larger angles. 

Fig.~\ref{fig:dd}(a) shows that when the number of scatterers increases, and hence the density of scatterers, the distribution of mutual scattering becomes statistically less dispersive. 
The mutual scattering at high density is then less sensitive to the internal configurations of the sample. 
While the density of the scatterers increases, the sample becomes increasingly opaque and the mutual scattering converges to its average values, that is, the blue and orange lines in Fig.~\ref{fig:statmet3}(b) converge, with a sine cardinal shape. 
Therefore, we associate the sine cardinal behaviour of the average value of mutual scattering with the external shape of the sample. 
At the same time, Fig.~\ref{fig:statmet3}(b) shows that the larger the angle $\gamma$, the wider the distribution of maximum and minimum mutual scattering, corresponding to blue and orange zones in Fig.~\ref{fig:statmet3}(b), are. 
Consequently, as the angle $\gamma$ increases, the mutual scattering is more influenced by the internal structure of the sample, and the sine cardinal shape of mutual scattering gradually becomes fainter. 
Summarizing this section, from a practical point of view, the mutual scattering at large angles holds information on the internal structure of the sample, and \textit{vice versa} the mutual scattering at small angles characterizes a sample's external shape.

\subsection{Asymmetric incident beams}
\begin{figure}[ht!]
\centering
\includegraphics[width=0.8\textwidth]{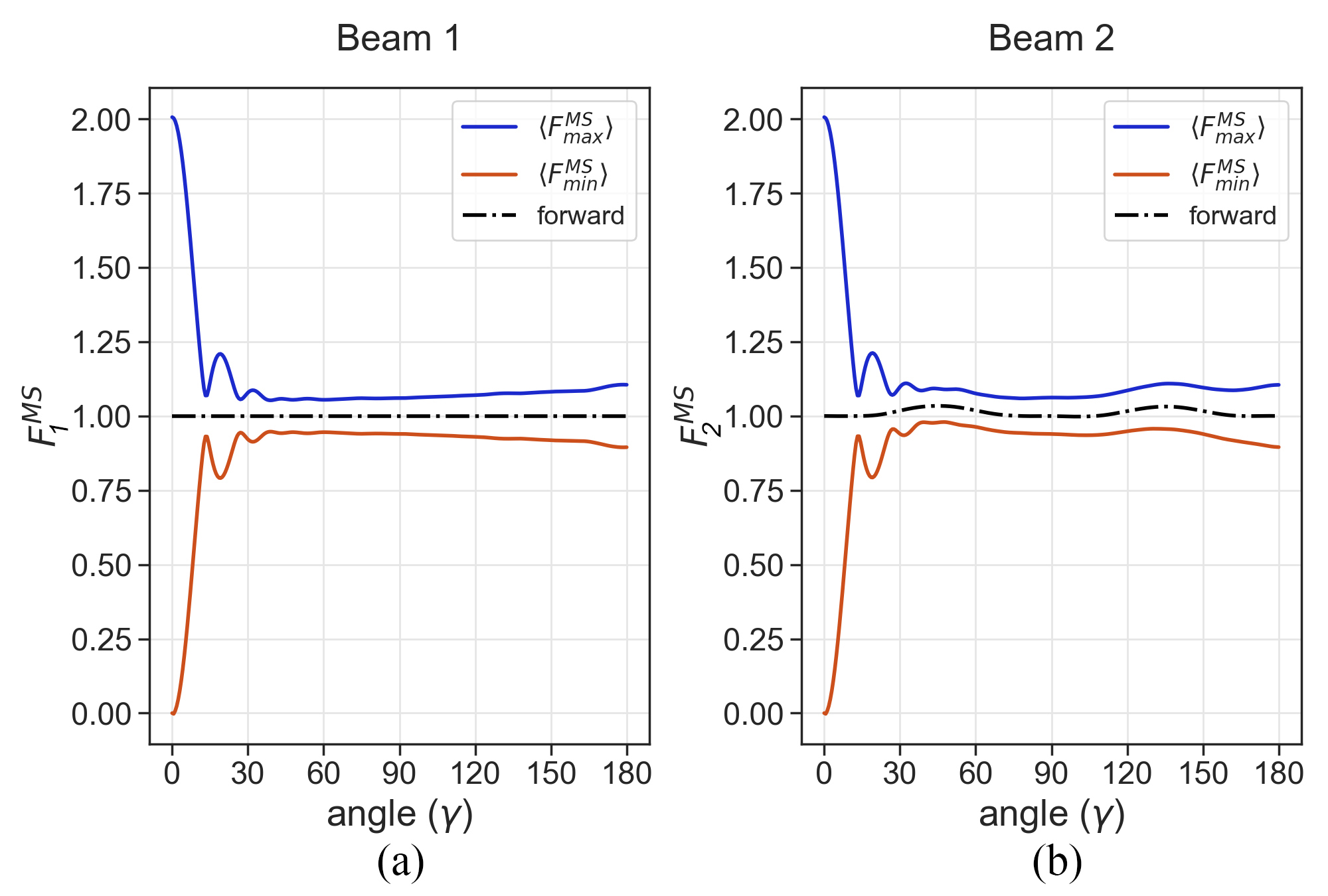}
\caption{\label{fig:statmet4}(a) Mutual scattering of the first beam $F^\text{MS}_1$ from two asymmetric incident beams with respect to the angle ($\gamma$) for $N_\text{dipole}= 250$ and $N_\text{realization}=1000$. (b) Mutual scattering of the second beam $F^\text{MS}_2$.}
\end{figure}

For the case of asymmetric incident beams, the incoming direction of the first beam $\mathbf{k}_{\text{in},1}$ is fixed. 
The angle of incidence of the first and second beams are no longer equal and are denoted as $\gamma_1=0\degree$ and $\gamma_1=\gamma$, respectively. 
As a result, the symmetry of mutual scattering between beam 1 and beam 2 no longer holds. 
For example, the difference between the mutual scattering of beam 1 ($F_1^\text{MS}$ detected at the sensor $\mathcal{A}$) and beam 2 ($F_2^\text{MS}$ detected at the sensor $\mathcal{B}$) is illustrated by Fig.~\ref{fig:statmet4}(a,b).

The most obvious difference between Fig.~\ref{fig:statmet4}(a) and Fig.~\ref{fig:statmet4}(b) is the symmetry of Fig.~\ref{fig:statmet4}(a). 
In particular, the average lines of the maximum and minimum mutual scattering are perfectly symmetrical about the black dash-dotted line (the forward scattering line $\langle F^\text{forward}_1(\gamma) \rangle$), which has a value of 1 at all angles $\gamma$. 
This is explained by the fact that the angle of incidence of the first beam is $\gamma_1=0\degree$. 
On the other hand, for beam 2 in Fig.~\ref{fig:statmet4}(b), the forward scattering line $\langle F^\text{forward}_2(\gamma) \rangle$  reach local maxima at $\gamma_2=45\degree$ and $\gamma_2=135\degree$. 

\section{Results II - Comparison between two-beam mutual scattering and one-beam techniques}
The goal of this section is to compare the two-beam mutual scattering and the traditional one-beam scattering.
If the second beam in Fig.~\ref{fig:geo1} is turned off and the incoming direction of the first beam $\mathbf{k}_{\text{in},1}$ is fixed at $0\degree$ ($\gamma_1=0\degree$), we have the conventional one-beam experiment.
Then, the sensor at $\mathcal{B}$ in Fig.~\ref{fig:geo1} only detects the current of the field scattered from direction $\mathbf{k}_{\text{in},1}$ into direction $\mathbf{k}_{\text{in},2}$. This scattered current can be considered as the angular ``speckle'' of the object~\cite{Goodman2020SpeckleEdition}, which is proportional to the differential cross-section from direction $\mathbf{k}_{\text{in},1}$ into direction $\mathbf{k}_{\text{in},2}$:
\begin{align}\label{eq:dcs}
    \dfrac{d\sigma}{d\Omega}(\gamma) \equiv \dfrac{d\sigma}{d\Omega}(\mathbf{k}_{\text{in},2},\mathbf{k}_{\text{in},1}) = \vert f(\mathbf{k}_{\text{in},2},\mathbf{k}_{\text{in},1}) \vert^2,
\end{align}
where $f(\mathbf{k}_{\text{in},2},\mathbf{k}_{\text{in},1})$ is the scattering amplitude of the object. On the other hand, the mutual scattering of the second beam $F^\text{MS}_{2}$ contains the interference between the incident field in the direction $\mathbf{k}_{\text{in},2}$ and the field scattered from direction $\mathbf{k}_{\text{in},1}$  into direction $\mathbf{k}_{\text{in},2}$.

\begin{figure}[ht!]
\centering
\includegraphics[width=0.62\textwidth]{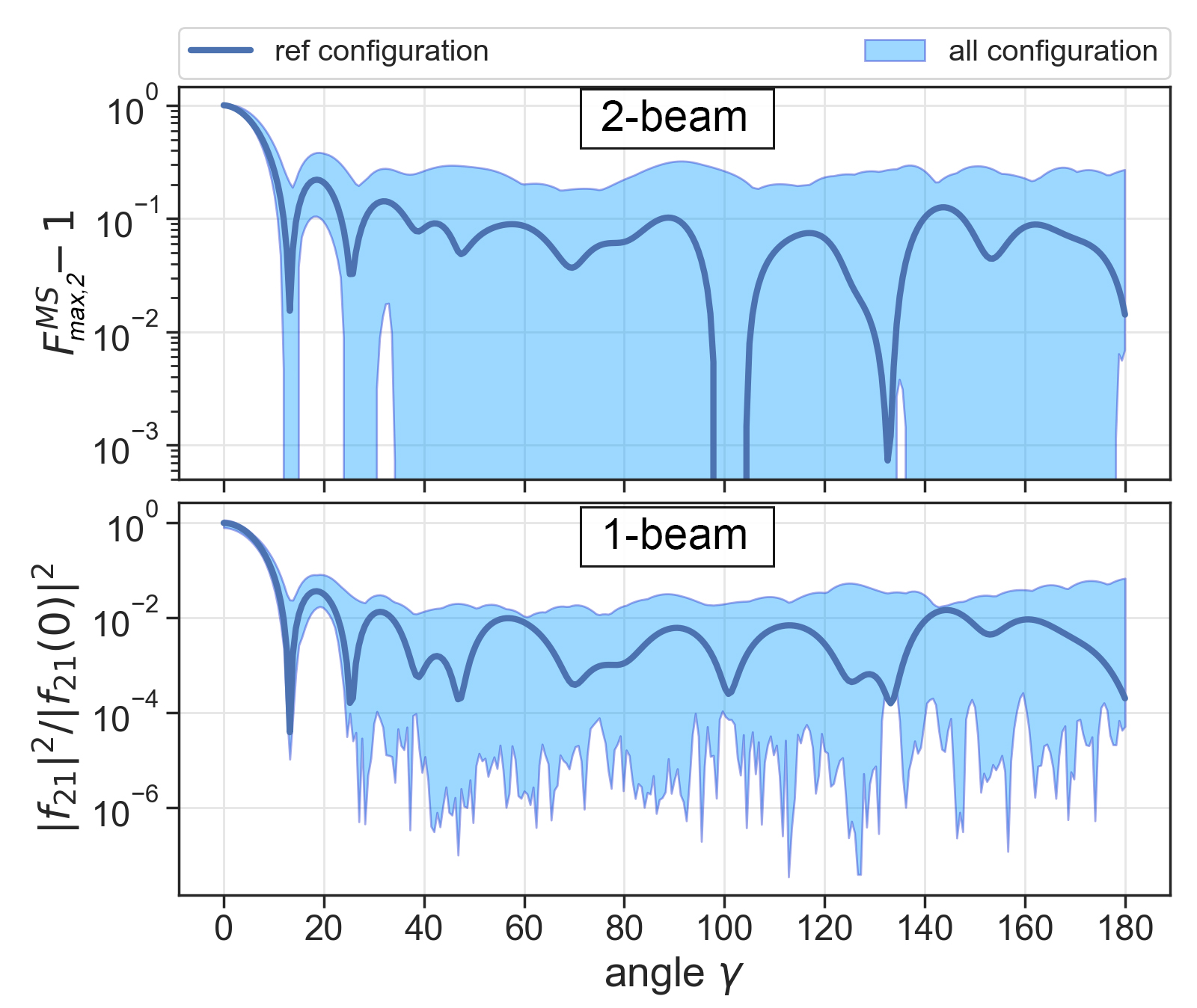}
\caption{\label{fig:2}
Comparison in a semi-log scale between (top) the maximum mutual scattering of the second beam $F^\text{MS}_{\text{max},2}$ in the case of asymmetric incidence and (bottom) the differential cross-section $\vert f_{21} \vert^2 \equiv \vert f(\mathbf{k}_2,\mathbf{k}_1) \vert^2$ with $N_\text{dipole}= 250$, and $N_\text{realization}=50$. 
The blue zones show how the maximum mutual scattering and the differential cross-section vary as the configurations of the dipoles are changed. 
}
\end{figure}

Fig.~\ref{fig:2}(top, bottom) shows the comparison between the one-beam differential cross-section and the maximum mutual scattering for a fixed ``reference'' configuration of 250 dipoles. 
It is worth noting that two reference blue lines in Fig.~\ref{fig:2}(top, bottom) tend to fluctuate up and down similarly at the same angles, but with different magnitudes. 
As is apparent in Fig.~\ref{fig:2}(bottom), when the angle $\gamma$ between the two beams increases, the differential cross-section $d\sigma/d\Omega$ decreases sharply to zero and varies in the range from $10^{-4}$ to $10^{-2}$. 
On the other hand, the mutual scattering in Fig.~\ref{fig:2}(top) experiences a milder variation as the value $F^{MS}_{max,2}-1$ of the reference configuration mostly fluctuates around $10^{-1}$. 
The 10 to 1000 times larger magnitude of mutual scattering compared to the differential cross-section is understood mathematically from the fact that the interference part of $F^\text{MS}_{2}$ is proportional to the imaginary part of the scattering amplitude ${\rm Im}\left[f(\mathbf{k}_{\text{in},1}, \mathbf{k}_{\text{in},2}) \right]$ (see Appendix \ref{appendix} for more details), while the differential cross-section Eq.~(\ref{eq:dcs}) gives the modulus squared of the scattering amplitude $\vert f(\mathbf{k}_{\text{in},1}, \mathbf{k}_{\text{in},2}) \vert^2$. 
In short, we consider the mutual scattering as a quantity that not only captures but also magnifies the behaviour of differential cross-section $d\sigma/d\Omega$, corresponding to the traditional speckle, and thus leads to a greater sensitivity that we will explore in the next section. 

\section{Results III - Sensing the position of a single displaced scatterer}
\label{sec:sen}
Knowing that two-beam mutual scattering is more sensitive than conventional one-beam scattering with respect to internal position variations of scatterers inside an opaque sample, we now turn to how the displacement of a single nanoparticle deep inside a sample is sensed with mutual scattering.

\subsection{Setup}
In brief, the procedure of our numerical setup is as follows:
\begin{enumerate}
    \item We start with a ``reference'' configuration: a selected dipole at position $\mathbf{r}_0$, and the other $(N_\text{dipole}-1)$ dipoles randomly distributed in the box, as illustrated in Fig.~\ref{fig:sche2}(b). 
    \item The selected dipole at $\mathbf{r}_0$ is displaced in the $x$-direction, as shown in Fig.~\ref{fig:sche2}(b), while the positions of all other ($N_\text{dipole}-1$) scatterers are fixed. 
    \item We calculate the difference in mutual scattering between the reference configuration and the new configuration obtained after the displacement of the red-sphere dipole. 
    \item We repeat the process with a new reference configuration where the selected dipole is still located at $\mathbf{r}_0$ but the other $(N_\text{dipole}-1)$ dipoles have randomly changed positions. 
    We compute the resulting statistic after $N_\text{realization}$ different configurations. 
\end{enumerate}

\begin{figure}[ht!]
\centering
\includegraphics[width=0.7\textwidth]{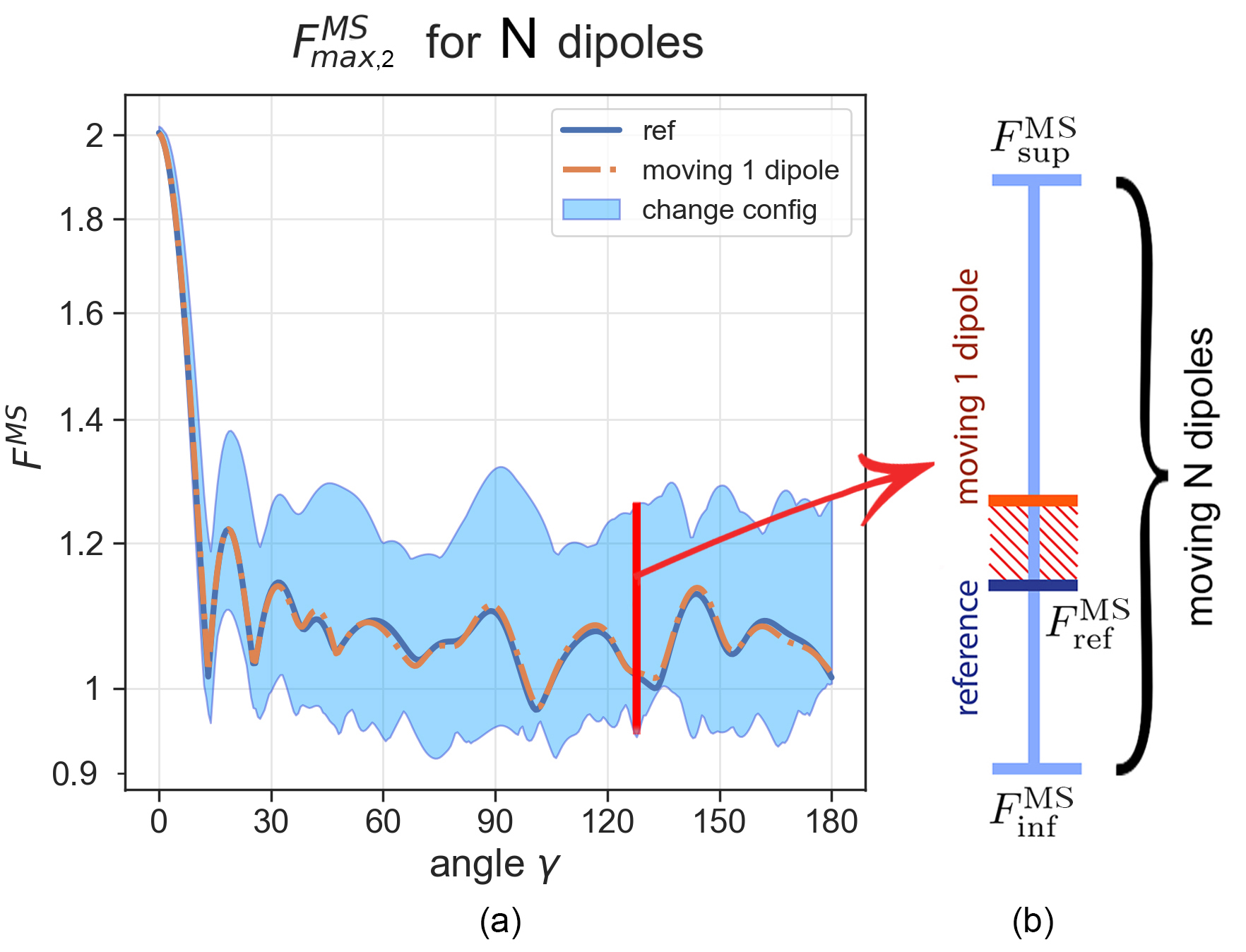}
\caption{\label{fig:sen0} Illustration of the susceptivity of mutual scattering for $N_\text{dipole}= 250$: 
(a) The solid blue line represents the reference maximum scattering before moving the selected dipole, i.e., the red sphere in Figure \ref{fig:sche2}(b). The dash orange line is the maximum scattering computed when the selected dipole has been displaced to the new position. 
The light blue zone covers all the values of maximum scattering for $N_\text{realization}=50$ realizations. 
(b) The susceptivity of mutual scattering (\ref{eq:eq1}) is defined as the ratio of the variation upon moving one scatterer (the orange zone) to the variation upon moving all $N_\text{dipole}$ scatterers (the light blue zone).}
\end{figure}

We quantify the variation of mutual scattering relative to the displacement of a single scatterer by a mathematical quantity called ``susceptivity'', which is defined based on Fig.~\ref{fig:sen0}(a,b). 
For example, the susceptivity of two-beam mutual scattering is defined as the ratio of variation upon moving one scatterer, i.e., the orange slashed zone in Fig.~\ref{fig:sen0}(b), to the variation upon the complete change of configuration of selection of scatterers when moving all $N_\text{dipole}$ scatterers, i.e., the light blue line in Fig.~\ref{fig:sen0}(b), by the following equation:
\begin{align}\label{eq:eq1}
\tilde{\delta} F^{\text{MS}} \coloneqq \dfrac{\left\vert F^\text{MS} -  F^\text{MS}_\text{ref}\right\vert}{F^\text{MS}_\text{sup}-F^\text{MS}_\text{inf}},
\end{align}
where $F^\text{MS}_\text{ref}$ stands for the reference configuration. 
After moving the selected dipole to a new position, the susceptivity of mutual scattering (compared to reference configuration $F^\text{MS}_\text{ref}$) is denoted by $\tilde{\delta} F^{\text{MS}}$. 
$F^\text{MS}_\text{sup}$ and $F^\text{MS}_\text{inf}$  refer to the limit superior and limit inferior which bound the variation of mutual scattering with respect to all the possible configurations of scatterers.
The susceptivity of 1-beam differential cross-section is also calculated in a similar way.

\subsection{Results and discussion}
\begin{figure}[ht!]
\centering
\includegraphics[width=0.95\textwidth]{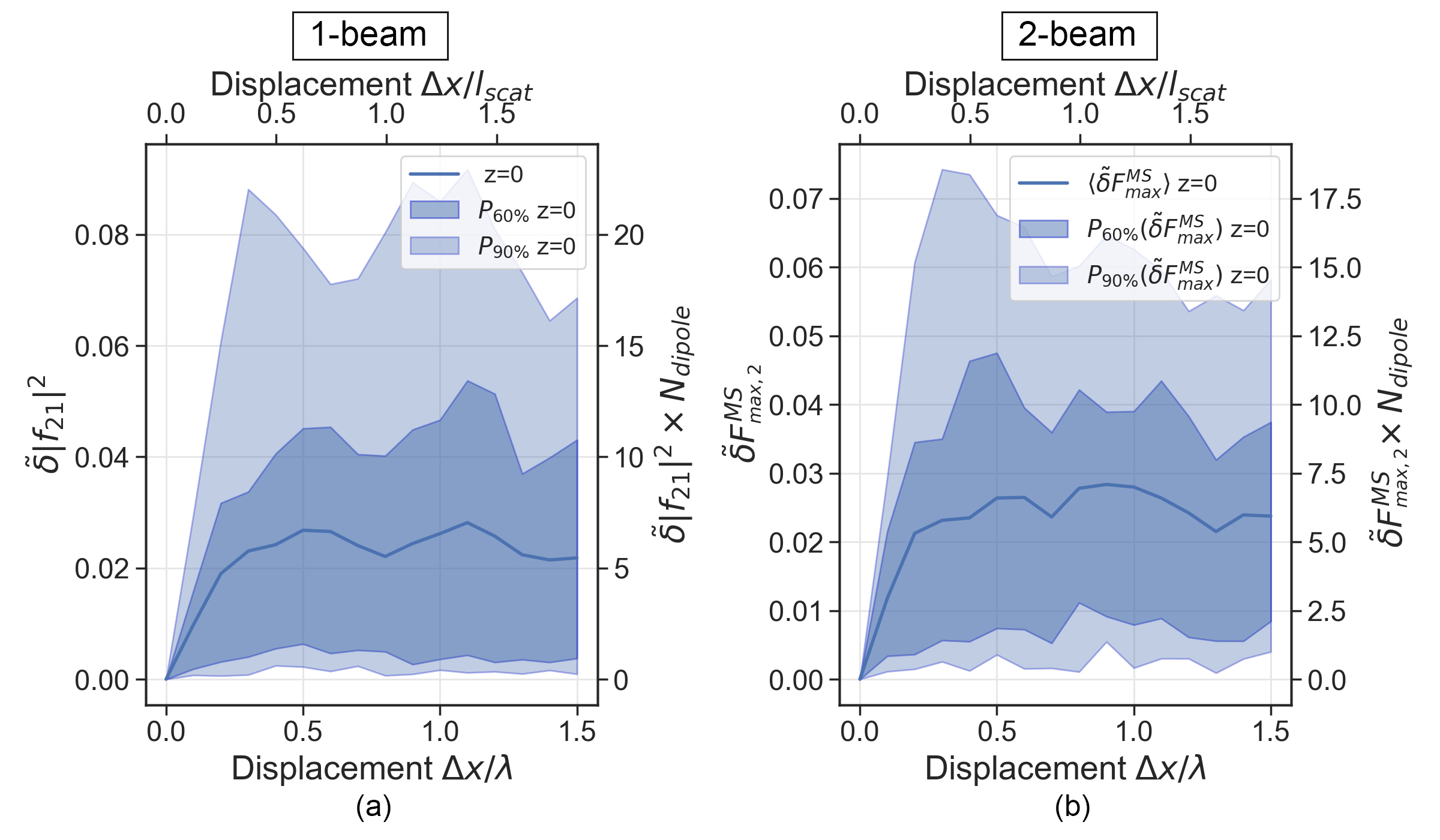}
\caption{\label{fig:sen1} Comparison between (a) the susceptivity of differential cross-section $\tilde{\delta}\vert f_{12} \vert^2$  and (b) the susceptivity of the maximum scattering of the second beam $\tilde{\delta} F^{\text{MS}}_{\text{max},2}$ with respect to the displacement in the $x$-direction of moving dipole, measured at angle $\gamma=90\degree$ for $N_\text{dipole}= 250$ and $N_\text{realization}=50$.   }
\end{figure}

We express the susceptivity of mutual scattering in Eq.~\ref{eq:eq1} as a function of displacement of a single dipole originally located at position $\mathbf{r}_0=(0,0,0)$, i.e., the centre of our box sample. 
The selected dipole is shifted along the $x$-axis. 
The statistics of the susceptivity of the maximum scattering of the second beam $\tilde{\delta} F^{\text{MS}}_{\text{max},2}$ and the differential cross-section $\tilde{\delta}\vert f_{21} \vert^2 \equiv \tilde{\delta}\vert f(\mathbf{k}_{\text{in},2},\mathbf{k}_{\text{in},1}) \vert^2$ with respect to the displacement $\Delta x$ at angle $\gamma=90\degree$ are shown in Fig.~\ref{fig:sen1}(a,b). 
Data are calculated with $N_\text{dipole}= 250$ dipoles, based on $N_\text{realization}=50$ different reference configurations of the location of dipoles.
The solid blue lines stand for the mean value of the susceptivity, while the dark and light blue zones are the 60-percentile and 90-percentile of the probability distribution of the susceptivity, respectively. 
The displacement value $\Delta x$ is normalized based on wavelength $\lambda$ and scattering mean free path $l_\text{scat}$. 

Let's take a closer look on the susceptivity of the maximum scattering of the second beam ($\tilde{\delta} F^{\text{MS}}_{\text{max},2}$)  in Fig.~\ref{fig:sen1}(b). 
The value of susceptivity tends to sharply increase in the range $\Delta x/\lambda \in [0,0.5]$, then reaches the asymptotic value and stays  around 0.025, over the range $\Delta x/\lambda \in [0.5,1.5]$. 
Intuitively, when changing one dipole out of $N_\text{dipole}$ dipoles, the susceptivity function is expected to change with a value of approximately the ratio $1/N_\text{dipole}$. 
In other words, we expect $\tilde{\delta} F^{\text{MS}}_{\text{max},2} \times N_\text{dipole} \approx 1$. 
However, as seen on the right twin $y$-axis in Fig.~\ref{fig:sen1}(b), the asymptotic value of mutual scattering at large displacement is on average 6 times the expected rate.

In practice, the transformation of ``one-beam speckle'', corresponding to the differential cross-section, as a function of displacement of a single scattering has never been experimentally measured because the magnitude of the differential cross-section at large angles is too small. 
However, we notice that the susceptivity of differential cross-section $\tilde{\delta} \vert f_{12} \vert^2$ in Fig.~\ref{fig:sen1}(a) behaves very similar to the susceptivity of the maximum scattering $\tilde{\delta} F^{\text{MS}}_{\text{max},2}$ in Fig.~\ref{fig:sen1}(b).
This allows us to study the dependence of speckles on the structure of the sample for the first time through the property of multi-beam mutual scattering.

After expressing the susceptivity of the maximum mutual scattering of the second beam $\tilde{\delta} F^{\text{MS}}_{\text{max},2}$ as a function of the displacement of the dipole from the centre of the sample, the next step is to find out how the susceptivity changes if the starting point of the moving scatterer is located elsewhere in the sample. 
In particular, let the original starting point of the moving (red) dipole be $\mathbf{r}_1=(0,0,z)$, where $z$ represents how deep the dipole lies inside of the sample relative to the incident surface.  
We choose 3 values of $z$:
\begin{itemize}
\item $z=-1.75$: The dipole is located very close to the incident surface. 
\item $z=0$: The dipole is at the center of the box.
\item $z=1.75$: The dipole is located very far from the incident surface and near the exit surface. 
\end{itemize}

\begin{figure}[ht!]
\centering
\includegraphics[width=.95\textwidth]{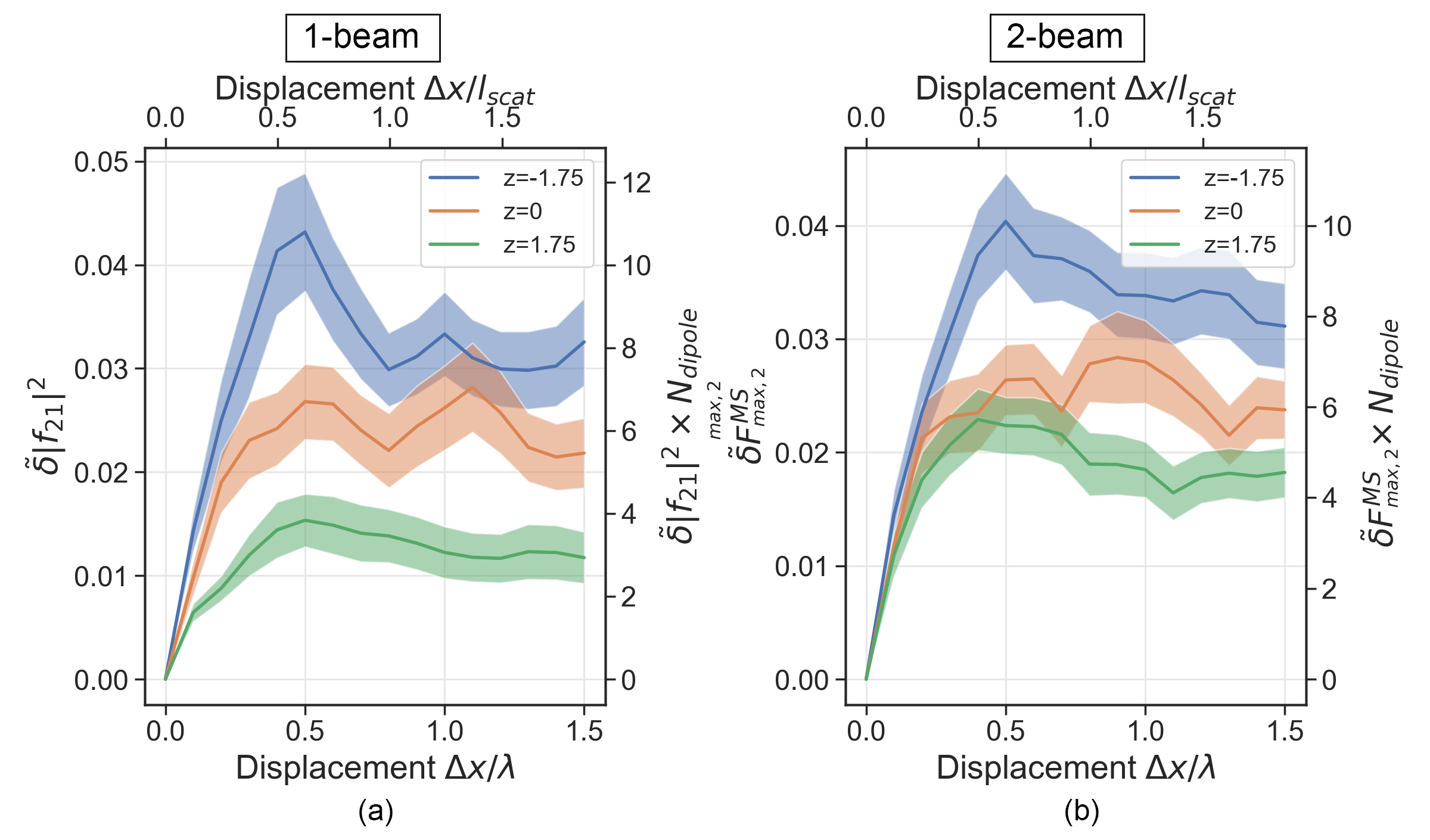}
\caption{\label{fig:sen5} Comparison between (a) the susceptivity of differential cross-section $\vert f_{21} \vert^2 \equiv \vert f(\mathbf{k}_2,\mathbf{k}_1) \vert^2$ and (b) the susceptivity of the maximum mutual scattering $\tilde{\delta} F^{\text{MS}}_{\text{max},2}$ of beam 2 in the case of asymmetric incidence  with respect to the displacement of the moving dipole at angle $\gamma=90\degree$ for $N_\text{dipole}= 250$ and $N_\text{realization}=50$. 
The figure depicts 3 different depths $z=-1.75, z = 0$,  and $z=1.75$ corresponding to 3 colors blue, orange, and green respectively. The solid lines stand for the mean value, while the colored zones represent the standard error of the mean. }
\end{figure}

We plot the susceptivity of the maximum mutual scattering of the second beam $\tilde{\delta} F^{\text{MS}}_{\text{max},2}$  with respect to the $x$-direction displacement of the moving dipole from $\mathbf{r}_1=(0,0,z)$ for all three values of $z$ in Fig.~\ref{fig:sen5}(b). 
Similar to Fig.~\ref{fig:sen1}(b), the susceptivity functions increase and then fluctuate around a fixed value, which depends on the original depth $z$ of the displaced dipole. 
We see that the blue line is always above the orange line, which is, in turn, above the green line. 
In other words, the susceptivity of the dipole located close to the incident surface ($z=-1.75$) fluctuates at a larger asymptote than the susceptivity of the dipole located far from the incident surface ($z=1.75$). 
In addition, the standard error of the mean of the susceptivity tends to decrease as the original location of the displaced dipole is further away from the incident surface.

There is also a surprising similarity between the susceptivity of the maximum mutual scattering in Fig.~\ref{fig:sen5}(b) and the differential cross-section in Fig.~\ref{fig:sen5}(a). 
This consolidates the fact that we can use mutual scattering as a substitute for differential cross-section, i.e., the modulus squared of scattering amplitude, in order to detect the location of a single scatterer. 

\begin{figure}[ht]
\centering
\includegraphics[width=0.65\textwidth]{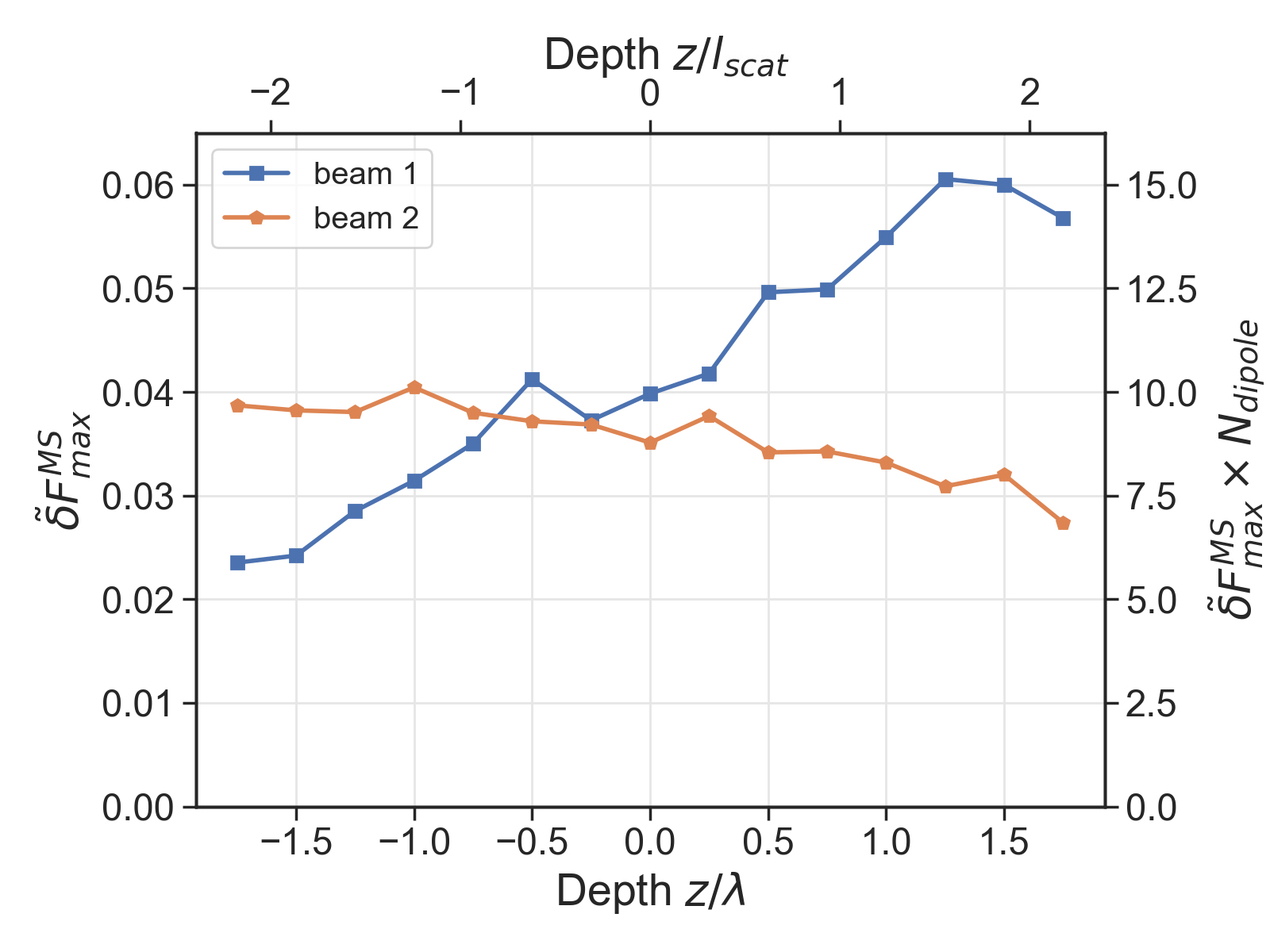}
\caption{\label{fig:senasym6a} Average of susceptivity function of the maximum mutual scattering $\tilde{\delta} F^{\text{MS}}_\text{max}$ of beam 1 (the blue line) and beam 2 (the orange line) with respect to the depth of the moving dipole. }
\end{figure}

The final conclusion of this paper is summarized by Fig.~\ref{fig:senasym6a}, which expresses the average of susceptivity function of the maximum mutual scattering of beam 1 $\langle \tilde{\delta} F^{\text{MS}}_{\text{max},1} \rangle$, i.e., the blue line, and beam 2 $\langle \tilde{\delta} F^{\text{MS}}_{\text{max},2} \rangle$,i.e., the orange line,  as functions of the depth of the displaced dipole in the case of asymmetric incidence. 
The result is computed for $N_\text{dipole}= 250$ and $N_\text{realization}=50$ at the displacement $\Delta x=1.5\lambda$ and averaged over all the angles $\gamma$. In agreement with Fig.~\ref{fig:sen5}(b), Fig.~\ref{fig:senasym6a} shows that the susceptivity of mutual scattering of beam 2 decreases as the depth $z$ (relative to the incident surface) of the displaced dipole in the x-direction increases. 
This property is understood through the scattering amplitude $f(\mathbf{k}_{\text{out}}, \mathbf{k}_{\text{in}})$ of multiple scatterers. 
For a fixed $\mathbf{k}_{\text{in},1}$, we conjecture that scattering amplitude $f(\mathbf{k}_{\text{in},2}, \mathbf{k}_{\text{in},1})$, i.e., the scattering strength from $\mathbf{k}_{\text{in},1}$ to $\mathbf{k}_{\text{in},2}$, will be more affected by scatterers located closer to the incident surface (in the direction  $\mathbf{k}_{\text{in},1}$) than ones located further away. 
On the other hand, the susceptivity of mutual scattering of beam~1 tends to grow with the depth $z$. This is explained as follows: 
\begin{enumerate}
\item Since scattering amplitude $f(\mathbf{k}_{\text{in},2}, \mathbf{k}_{\text{in},1})$ is more affected by the scatterers located closer to the incident surface (in the direction $\mathbf{k}_{\text{in},1}$), by the reciprocity of light, for a fixed $\mathbf{k}_{\text{in},1}$, the scattering amplitude $f(-\mathbf{k}_{\text{in},1},-\mathbf{k}_{\text{in},2})$ (the scattering strength from $-\mathbf{k}_{\text{in},2}$ to $-\mathbf{k}_{\text{in},1}$) will be more affected by scatterers located further from the incident surface (in the direction  $-\mathbf{k}_{\text{in},1}$).
\item Since $f(-\mathbf{k}_{\text{in},1},-\mathbf{k}_{\text{in},2})$ is more affected by scatterers located further from the incident surface (in the direction  $-\mathbf{k}_{\text{in},1}$), by the symmetry of our box sample, for a fixed $\mathbf{k}_{\text{in},1}$, the scattering amplitude $f(\mathbf{k}_{\text{in},1},\mathbf{k}_{\text{in},2})$ will be more affected by scatterers located further from the incident surface (in the direction  $\mathbf{k}_{\text{in},1}$).
\end{enumerate}

All these statistical properties are used to determine the location of the moving dipole, or even further, detect motions inside opaque objects. 

\section{Discussion} 
\label{sec:discussion}

The superior sensitivity of the two-beam mutual scattering over the one-beam differential cross-section with respect to the subtle internal variations of the sample opens up a huge potential to explore structural data which are very difficult to extract from traditional one-beam speckle. 
One of the most interesting applications is in imaging of the movement of tissues and body fluids, which until now still relies on the Doppler effect in ultrasonography~\cite{Srivastav2019IEEE}. 
The complex structure of biological tissues requires expensive high-tech tools to bypass the multiple scattering problem. 
Thus, if possible, statistically detecting the movement of tissues and body fluids through mutual scattering could open up new possibilities for optical applications in medicine.

Verifying the current theoretical predictions in experiments will guide the next development of the research on the applications of mutual scattering. 
Our near-future goal is to measure the statistical difference in the susceptivity of mutual scattering with respect to the change of displacement of the internal fraction of opaque media in the upcoming experiments.  

One of the further research directions of this topic is to fully develop a method and schematic diagram based on mutual scattering for extracting a complete profile of the complex scattering amplitude of a given sample for all incoming and outgoing wave vectors. Since mutual scattering allows us to extract the imaginary part of the scattering amplitude ${\rm Im}\left[f(\mathbf{k}_{\text{out}}, \mathbf{k}_{\text{in}}) \right]$, it is possible to reconstruct the full profile of the full complex profile of scattering amplitude $f(\mathbf{k}_{\text{out}}, \mathbf{k}_{\text{in}})$ of the object for all incoming and outgoing directions.

Finally, it is worth investigating the potential of speckle correlation of ``two-beam speckle patterns'', which is used commonly for single-beam techniques~\cite{Bertolotti2012Nature,JaureguiSanchez2022NatureCommunications}, as another way to extract more information about the shape and movement of the object.

\section{Conclusion}
\label{sec5}
In this paper, we discuss potential applications of mutual scattering in the study of the internal structure of matter.
In particular, mutual scattering from incoming beams, which possesses the same order of magnitude as the scattering amplitude, is easily measured with much higher accuracy than the scattering current from one beam, which is proportional to the modulus square of the scattering amplitude. 
Adding an extra beam (on top of traditional one-beam techniques) and measuring ``two-beam speckle patterns'' allows us to more precisely extract statistical data regarding the internal structure of objects, which is usually not properly appreciated and is considered as noise to be eliminated. 

Moreover, we demonstrate through a numerical example how information about the depth of a displaced dipole in an opaque box is obtained through the susceptivity of mutual scattering. 
In detail, the optical characteristics of the boxed sample are simulated by the multiple scattering problem of $N_\text{dipole}$ scatterers. 
The calculation results show that, at different depths, the susceptivity function of mutual scattering increases with displacement distance and fluctuates around different values. 
Studying these statistical data shows that the susceptivity of mutual scattering of beam 2 (or beam 1) tends to decline (or increase) as the depth of the displaced dipole increases, which in turn reveals the location of the moving dipole.

\section{Appendix - Theory}\label{appendix}
\subsection{Scattering with a single incoming wave}
We will limit ourselves to scalar waves, for mathematical convenience. 
In scattering theory, when scalar light is scattered from an object, the scalar field is partitioned into an unperturbed part and a scattered part:
\begin{align}
    \psi = \psi_\text{in}+ \psi_\text{scat},
\end{align}
wherein the case of single incident plane wave, the unperturbed part (the incident part) is given as follows:
\begin{align}
    \psi_{\text{in}}(\mathbf{r},\mathbf{k}_\text{in}) & = A \exp\left( i\mathbf{k}_\text{in} \cdot \mathbf{r} -i\omega t +i\phi \right). 
\end{align}
The real-valued $A$ represents the amplitude, $\omega$ and $\phi$ are the frequency and the phase of the incoming plane wave.
$\mathbf{k}_\text{in} = (\omega/c)  \hat{ \mathbf{k}}_\text{in} \equiv k \hat{ \mathbf{k}}_\text{in}$ is
the incoming wave vector. 
Then, the amplitude of the total wave in the far-field is given by:
\begin{align}
\lim_{r\to \infty} \psi(\mathbf{r}) & = \lim_{r\to \infty} \left( \psi_\text{in} + \psi_\text{scat} \right)
\nonumber \\ & =  A \exp\left( i\mathbf{k}_{\text{in}} \cdot \mathbf{r} -i\omega t +i\phi \right)  + \dfrac{A}{r} f\left(\mathbf{r},\mathbf{k}_{\text{in}}\right) \exp\left( ikr -i\omega t +i\phi \right),
\end{align}
where the scattering amplitude $f\left(\mathbf{k}_{\text{out}},\mathbf{k}_{\text{in}}\right)$ is the scattering strength from incoming direction $\mathbf{k}_{\text{in}}$ to outgoing direction $\mathbf{k}_{\text{out}}$, and $r=\vert\mathbf{r}\vert$. The scattering amplitude, as shown in Section \ref{sec:static} and Section \ref{sec:sen}, is an essential ingredient in extracting the internal structure of the sample.  

We recall that, experimentally, the quantity measured at far-field detectors is usually the current $\mathbf{J}$  of the scalar wave instead of the amplitude of the wave~\cite{Lagendijk1996PhyRep}:
\begin{align}
    \mathbf{J} \equiv -{\rm Re}\left[ (\partial_t \psi)^\ast \nabla \psi \right],
\end{align}
where we express the real and imaginary part of a complex value $x$ as $\rm Re[x]$ and $\rm Im[x]$, respectively. Then, the extinction of the wave is given by the interference of the incoming beams and the scattered beams:
\begin{align}
\mathbf{J}_\text{ext} = -{\rm Re}[(\partial_t\psi_\text{in})^\ast \nabla \psi_\text{scat}]-{\rm Re}[(\partial_t\psi_\text{scat})^\ast \nabla \psi_\text{in}].     
\end{align}

In the one-incoming-beam case, the observed scattering current measured at the direction $\mathbf{r}$ is given as follows:
\begin{align}\label{eq:scat}
\lim_{r\to \infty} \mathbf{J}_{\text{scat}} (\mathbf{r}, \mathbf{k}_{\text{in}}) = & \dfrac{\omega^2}{r^2c}A^2 \left\vert f(\mathbf{r}, \mathbf{k}_{\text{in}}) \right\vert^2 .  
\end{align}


\subsection{Scattering with multiple waves}\label{sec:Appendix_multiple}
The concept of mutual scattering occurs when there are multiple incoming waves. 
We focus on the scenario of two incoming plane waves:
\begin{align}
\psi_{\text{in}}(\mathbf{r},\mathbf{k}_\text{in}) & = \psi_{\text{in},1}(\mathbf{r},\mathbf{k}_{\text{in},1}) + \psi_{\text{in},2}(\mathbf{r},\mathbf{k}_{\text{in},2}) \nonumber\\
& = A_1 \exp\left( i\mathbf{k}_{\text{in},1} \cdot \mathbf{r} -i\omega t +i\phi_1 \right) + A_2 \exp\left( i\mathbf{k}_{\text{in},2} \cdot \mathbf{r} -i\omega t +i\phi_2 \right),
\end{align}
where we assume that the two beams have the frequency $\omega$. Let $A_1$ (or $A_2$) stand for the amplitude, $\mathbf{k}_{\text{in},1}$ (or $\mathbf{k}_{\text{in},2}$) the incoming direction and $\phi_1$ (or $\phi_2$) the phase of the first (or second) incoming plane wave. 

The amplitude of the waves at far-field for the case of two beams is shown as below:
\begin{align}
\lim_{r\to \infty} \psi(\mathbf{r}) = & \lim_{r\to \infty} \left( \psi_\text{in} + \psi_\text{scat} \right)
\nonumber \\ = & \, A_1 \exp\left( i\mathbf{k}_{\text{in},1} \cdot \mathbf{r} -i\omega t +i\phi_1 \right)  + \dfrac{A_1}{r} f\left(\mathbf{r},\mathbf{k}_{\text{in},1}\right) \exp\left(ikr -i\omega t +i\phi_2 \right) \nonumber \\
& + A_2 \exp\left( i\mathbf{k}_{\text{in},2} \cdot \mathbf{r} -i\omega t +i\phi_1 \right)  + \dfrac{A_2}{r} f\left(\mathbf{r},\mathbf{k}_{\text{in},2}\right) \exp\left(ikr -i\omega t +i\phi_2 \right).
\end{align}

The current of self-extinction of beam 1 along the forward scattering direction is given by:
\begin{align}
\lim_{r\to \infty} \mathbf{J}_{\text{in},1} (\mathbf{r},\mathbf{k}_{\text{in},1})= & -\dfrac{2\omega}{r^2}A^2{\rm Im}\left[f(\mathbf{k}_{\text{in},1}, \mathbf{k}_{\text{in},1})  \right]
\delta\left\{1-\cos(\mathbf{r},\mathbf{k}_{\text{in},1})\right\},
\end{align}
where $\cos(\mathbf{r},\mathbf{k}_{\text{in},1})$ is the trigonometric function of an angle between two vectors $\mathbf{r}$ and $\mathbf{k}_{\text{in},1}$.
$\delta\left\{1-\cos(\mathbf{r},\mathbf{k}_{\text{in},1})\right\}$ is the Dirac delta function, whose value is zero everywhere except at $1-\cos(\mathbf{r},\mathbf{k}_{\text{in},1})=0$, and whose integral over all values of $\cos(\mathbf{r},\mathbf{k}_{\text{in},1})$  is equal to one.

Then, the current of extinction detected along the direction $\mathbf{k}_{\text{in},1}$ consists of the incoming current of the first beam and the extinction of the field scattered from the direction $\mathbf{k}_{\text{in},2}$ into $\mathbf{k}_{\text{in},1}$:
\begin{align}\label{eq:Jext1}
\lim_{r\to \infty} \mathbf{J}_{\text{ext},1} (\mathbf{k}_{\text{in},2},\mathbf{k}_{\text{in},1}) = & -\dfrac{2\omega}{r^2}A^2_1 {\rm Im}\left[f(\mathbf{k}_{\text{in},1}, \mathbf{k}_{\text{in},1})  \right] \delta\left\{1-\cos(\mathbf{r},\mathbf{k}_{\text{in},1})\right\} \nonumber\\ 
& -\dfrac{2\omega}{r^2}A_1 A_2 {\rm Im} \left[f(\mathbf{k}_{\text{in},2}, \mathbf{k}_{\text{in},1}) e^{i(\phi_2-\phi_1)} \right] \delta\left\{1-\cos(\mathbf{r},\mathbf{k}_{\text{in},1})\right\}.
\end{align}

We note that the value of  $\lim_{r\to \infty} \mathbf{J}_{\text{ext},1}$ depends on the angle between two vectors $\mathbf{k}_{\text{in},1}$ and $\mathbf{k}_{\text{in},2}$. 
Let $\gamma$ denote this angle between $\mathbf{k}_{\text{in},1}$ and $\mathbf{k}_{\text{in},2}$, we express total extinction current of beam 1 at far-field as a function of $\gamma$:  $\lim_{r\to \infty} \mathbf{J}_{\text{ext},1}(\gamma)$. 
Similarly, we express the self-extinction $\lim_{r\to \infty} \mathbf{J}_{\text{in},1}(\gamma)$ as a function of $\gamma$.

\subsection{Mutual  scattering}
We define the normalized mutual  scattering of beam $i$, $i\in\{1,2\}$, as follows:
\begin{align}\label{eq:me1}
F^\text{MS}_i(\gamma) = \dfrac{\lim_{r\to \infty} \mathbf{J}_{\text{ext},i}(\gamma)}{\lim_{r\to \infty} \mathbf{J}_{\text{in},i}(\gamma=0)} \:,
\end{align}
where the denominator part is added to normalize the value of mutual  scattering. In fact, at $\gamma=0$, by tuning the phase difference $(\phi_1-\phi_2)$, the mutual  scattering  $F^\text{MS}_i(\gamma=0)$ has a maximum value of 2 and a minimum value of 0 (see for instance Fig.~\ref{fig:statmet3}). 

The self-extinction of beam $i$ containing only the forward scattering, denoted by $F^\text{forward}_{i}(\gamma)$, is normalized  by the following equation:
\begin{align}
F^\text{forward}_{i}(\gamma) \equiv \dfrac{\lim_{r\to \infty} \mathbf{J}_{\text{in},i}(\gamma)}{\lim_{r\to \infty} \mathbf{J}_{\text{in},i}(\gamma=0)} \; .  
\end{align}

The normalized mutual  scattering  of both two beams $F^\text{MS}_{12}(\gamma)$ is given by:
\begin{align}\label{eq:me2}
F^\text{MS}_{12}(\gamma) = \dfrac{\lim_{r\to \infty} [\mathbf{J}_{\text{ext},1}(\gamma) +\mathbf{J}_{\text{ext},2}(\gamma)]}{\lim_{r\to \infty} [\mathbf{J}_{\text{in},1}(\gamma=0)+ \mathbf{J}_{\text{in},2}(\gamma=0)]} \; .  
\end{align}
Equations~(\ref{eq:me1}) and (\ref{eq:me2}) will be used throughout this paper to calculate the mutual  scattering.
Finally, the forward-scattering (self-extinction) of both beams is given by
\begin{align}
F^\text{forward}_{12}(\gamma) = \dfrac{\lim_{r\to \infty} (\mathbf{J}_{\text{in},1}+\mathbf{J}_{\text{in},2})}{\lim_{r\to \infty} [\mathbf{J}_{\text{in},1}(\gamma=0)+ \mathbf{J}_{\text{in},2}(\gamma=0)]}.
\end{align}

\subsection{T-matrix for multiple scattering problem}
\label{sub:Tmatrix}
Given a complex scattering medium consisting of $N_\text{dipole}$ scatterers, it is important to notice that calculating the T-matrix of the whole sample requires summing all the scattering events order by order, up to infinite order, from a collection of all $N_\text{dipole}$ scatterers.  
The scattering of each point scatterer is, in turn, characterized by its single particle t-matrix. We recall the t-matrix for a single-point dipole (see~\cite{DeVries1998RevModPhy} for more details) as follows:
\begin{align}
t(\omega) =  -\dfrac{4\pi c}{\omega_0 Q} \dfrac{\omega_0^2}{\omega_0^2-\omega^2 -i\frac{\omega^3}{Q\omega_0}}, 
\end{align}
where $\omega_0$ is the resonance frequency and $Q$ is the quality factor of the resonance. 
For simplicity, the frequency $\omega$ is normalized to wavelength $\lambda=1$. The resonance frequency is set very close to the frequency of light $\omega_0= 1.0001\omega$. The speed of light is set to 1, $c=1$, and the quality factor is chosen to be 10, $Q=10$.

The polarizability $\alpha(\omega)$ in Fig.~\ref{fig:sche2} is obtained from the t-matrix of single point scatterer
\begin{align}
    \alpha(\omega) = -t(\omega) \dfrac{c^2}{\omega^2}.
\end{align}
From the t-matrix of single point scatterer, the full T-matrix  $T(\mathbf{k}_{\text{out}}, \mathbf{k}_{\text{in}})$ of the sample is obtained by inversion of a matrix $N_\text{dipole} \times N_\text{dipole}$, where calculation details are given in~\cite{Lagendijk2020EPL}. 



\begin{backmatter}
\bmsection{Funding}
This work is supported by the NWO-TTW program P15-36 “Free-Form Scattering Optics” (FFSO) and the MESA+ Institute section Applied Nanophotonics (ANP). 

\bmsection{Acknowledgments}
It is a great pleasure to thank Lars Corbijn van Willenswaard, and Alfredo Rates for their helpful discussions. 
AL and WLV are grateful to the staff of the Institut Langevin (Paris) for hospitality during their recent visits. 

\bmsection{Disclosures}
The authors declare no conflicts of interest. 

\bmsection{Data availability} 
Data underlying the results presented in this paper are available in Ref.~\cite{Truong2022dataset}. 
\end{backmatter}

\bibliography{ref_mut_scat_position}

\end{document}